\documentclass[notitlepage,
bibnotes,
aps,prb,10pt, 
twocolumn,
showpacs,preprintnumbers,superscriptaddress,
amsmath,amssymb,floatfix]{revtex4-1}
\usepackage{graphicx}
\usepackage{dsfont}
\usepackage{dcolumn}
\usepackage{bm}
\usepackage{color}
\usepackage{xcolor}
\usepackage{url}
\usepackage{tabularx}
\usepackage{array}
\usepackage{booktabs}
\usepackage{comment}
\usepackage{hyperref}
\usepackage{footnote}
\usepackage{lipsum}
\usepackage[normalem]{ulem}

\renewcommand{\onlinecite}[1]{ref.~\nocite{#1}\citenum{#1}} 
\usepackage{titlesec}
\newcommand*{\justifyheading}{\raggedright}
\titleformat{\section}
  {\normalfont\bfseries\justifyheading}{\thesection}{1em}{}  
\usepackage[singlelinecheck=false, justification=RaggedRight, format=plain]{caption}
\DeclareCaptionLabelSeparator{bar}{ $\boldsymbol{\mid}$ }
\newcommand{\thesupplementbibliography}{\thebibliography}
\usepackage{etoolbox}
\makeatletter
\apptocmd{\thesupplementbibliography}{\global\c@NAT@ctr 30\relax}{}{}
\makeatother
\begin{document}

\title{ 
Experimental observation of magnetic bobbers for a new concept of~magnetic~solid-state~memory
}

\author{Fengshan~Zheng}
\affiliation{Ernst Ruska-Centre for Microscopy and Spectroscopy with Electrons and Peter Gr\"unberg Institute, Forschungszentrum J\"ulich, 52425 J\"ulich, Germany}

\author{Filipp~N.~Rybakov}
\affiliation{KTH Royal Institute of Technology, Stockholm, SE-10691 Sweden}
\affiliation{M.N. Miheev Institute of Metal Physics of Ural Branch of Russian Academy of Sciences, Ekaterinburg 620990, Russia}
\affiliation{Ural Federal University, Ekaterinburg 620002, Russia}

\author{Aleksandr~B.~Borisov}
\affiliation{M.N. Miheev Institute of Metal Physics of Ural Branch of Russian Academy of Sciences, Ekaterinburg 620990, Russia}
\affiliation{Ural Federal University, Ekaterinburg 620002, Russia}

 \author{Dongsheng~Song}
 \affiliation{National Center for Electron Microscopy in Beijing, School of Materials Science and Engineering, Tsinghua University, Beijing 100084, China}

 \author{Shasha~Wang}
 \affiliation{The Anhui Key Laboratory of Condensed Matter Physics at Extreme Conditions, High Magnetic Field Laboratory, Chinese Academy of Science (CAS), Hefei, Anhui Province 230031, China}
 \affiliation{Collaborative Innovation Center of Advanced Microstructures, Nanjing University, Jiangsu Province 210093, China}
 
\author{Zi-An~Li}
 \affiliation{Institute of Physics, Chinese Academy of Sciences, 100190 Beijing, China}

\author{Haifeng~Du}
 \email{duhf@hmf.ac.cn}
 \affiliation{The Anhui Key Laboratory of Condensed Matter Physics at Extreme Conditions, High Magnetic Field Laboratory, Chinese Academy of Science (CAS), Hefei, Anhui Province 230031, China}
 \affiliation{Collaborative Innovation Center of Advanced Microstructures, Nanjing University, Jiangsu Province 210093, China}
 
\author{Nikolai~S.~Kiselev}
 \email{n.kiselev@fz-juelich.de}
 \affiliation{Peter Gr\"unberg Institute and Institute for Advanced Simulation, Forschungszentrum J\"ulich and JARA, 52425 J\"ulich, Germany}

\author{Jan~Caron}
\author{Andr\'as~Kov\'acs}

\affiliation{Ernst Ruska-Centre for Microscopy and Spectroscopy with Electrons and Peter Gr\"unberg Institute, Forschungszentrum J\"ulich, 52425 J\"ulich, Germany}

\author{Mingliang~Tian}
 \affiliation{The Anhui Key Laboratory of Condensed Matter Physics at Extreme Conditions, High Magnetic Field Laboratory, Chinese Academy of Science (CAS), Hefei, Anhui Province 230031, China}
 \affiliation{Collaborative Innovation Center of Advanced Microstructures, Nanjing University, Jiangsu Province 210093, China}

\author{Yuheng~Zhang}
 \affiliation{The Anhui Key Laboratory of Condensed Matter Physics at Extreme Conditions, High Magnetic Field Laboratory, Chinese Academy of Science (CAS), Hefei, Anhui Province 230031, China}
 \affiliation{Collaborative Innovation Center of Advanced Microstructures, Nanjing University, Jiangsu Province 210093, China}
 
 \author{Stefan~Bl\"ugel}
 \affiliation{Peter Gr\"unberg Institute and Institute for Advanced Simulation, Forschungszentrum J\"ulich and JARA, 52425 J\"ulich, Germany}

\author{Rafal~E.~Dunin-Borkowski}
\affiliation{Ernst Ruska-Centre for Microscopy and Spectroscopy with Electrons and Peter Gr\"unberg Institute, Forschungszentrum J\"ulich, 52425 J\"ulich, Germany}

\date{May 8, 2017}

\maketitle

\textbf{The use of chiral skyrmions~\cite{Bogdanov_89,Ivanov_90,Kiselev11}, which are nanoscale vortex-like spin textures, as movable data bit carriers forms the basis of a recently proposed concept for magnetic solid-state memory~\cite{Fert_13}. 
In this concept, skyrmions are considered to be unique localized spin textures~\cite{Bogdanov_89,Koshibae}, which are used to encode data through the quantization of different distances between identical skyrmions on a guiding nanostripe.
However, the conservation of distances between highly mobile and interacting skyrmions is difficult to implement in practice. 
Here, we report the direct observation of another type of theoretically-predicted localized magnetic state, which is referred to as a chiral bobber (ChB)~\cite{Rybakov_15}, using quantitative off-axis electron holography.
We show that ChBs can coexist together with skyrmions.
Our results suggest a novel approach for data encoding, whereby a stream of binary data representing a sequence of ones and zeros can be encoded via a sequence of skyrmions and bobbers.
The need to maintain defined distances between data bit carriers is then not required. 
The proposed concept of data encoding promises to expedite the realization of a new generation of magnetic solid-state memory.
}

Although magnetic hard disk drives show great reliability and are some of the most in-demand devices on the market, the ultimate data density and operating speed are limited by superparamagnetic effects~\cite{Weller_99} and by the presence of energy-consuming mechanical components, such as engines and actuators. 
Therefore, alternative approaches for solid state magnetic memory devices that have no movable mechanical parts have been proposed.
Such approaches should achieve higher performance, data capacity, reliability and energy efficiency.

\begin{figure*}
\centering
\includegraphics[width=18cm]{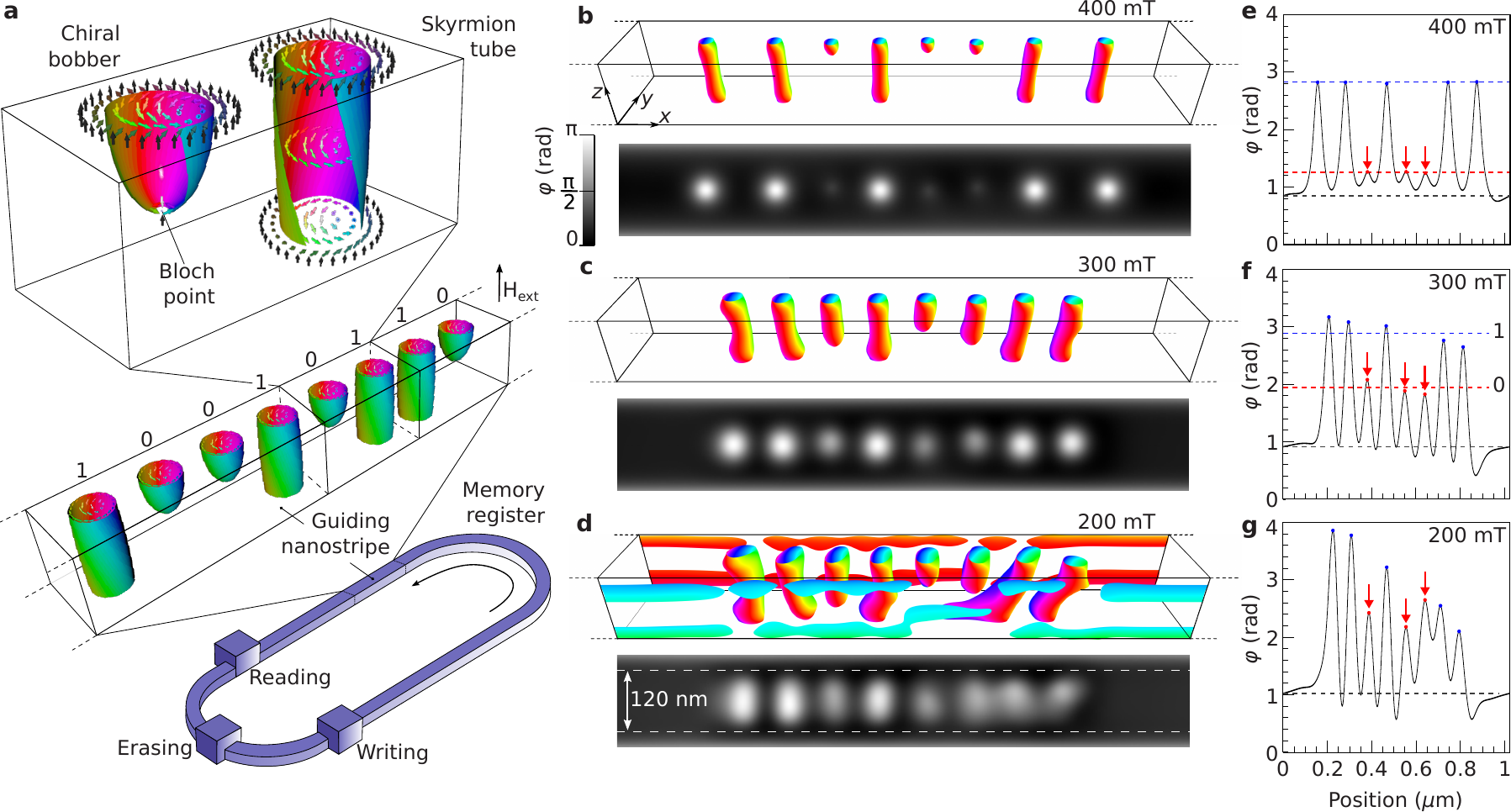}
\caption{\small \textbf{New concept for magnetic solid-state memory.} 
\textbf{a},~Schematic representation of an encoded data stream in a nanostripe, which takes the form of a closed track containing a chain of alternating magnetic skyrmions and chiral bobbers, which correspond to  ``1'' and ``0'' bits.
The actions of writing, reading and erasing the units are performed at different positions along the guiding track.
Here and below, the colors of vectors and isosurfaces for $m_z\!=\!0$ are defined by the in-plane component of the magnetisation according to a standard color wheel.
\textbf{b-d},~Results of micromagnetic simulations performed for a chain of skyrmion tubes (SkTs) and chiral bobbers (ChBs) in a 1024~$\times$~164~$\times$~128~nm FeGe nanostripe. 
In each case, the top panel shows isosurfaces for $m_z\!=\!0$, while the bottom panel shows corresponding theoretical predictions of the phase shift of an electron beam measured using off-axis electron holography for different values of the externally applied magnetic field along $\mathbf{e}_z$.
\textbf{e-g},~Calculated magnetic-field-induced phase shift averaged over a band of width 120~nm, as indicated in \textbf{d}. 
The blue, red and black dashed lines in \textbf{e},~\textbf{f} correspond to the average signal of a SkT, a ChB and the nearly homogeneous state in between them, respectively. The maximum values that correspond to ChBs are indicated by red arrows.
}
\label{Fig-intro}
\end{figure*}

One of the most promising candidates for a new solid-state magnetic memory device is based on the concept of racetrack memory (RM)~\cite{Parkin_15}, in which the role of data bit carriers is played by either (i) domain walls, which are small transition regions between domains whose magnetization typically points in opposite directions~\cite{Parkin_15} or (ii) chiral skyrmions ~\cite{Bogdanov_89, Ivanov_90, Kiselev11}, which possess topologically protected stability and can be moved using currents that are several orders of magnitude lower than those required for magnetic domain wall motion~\cite{Fert_13}.
Skyrmion-based RM is currently considered to be the most promising approach. 

Chiral magnetic skyrmions appear in ferromagnetic crystals that have a particular symmetry and strong spin orbit coupling, giving rise to the Dzyaloshinskii-Moriya interaction (DMI)~\cite{Dzyaloshinskii,Moriya}. 
In chiral magnets, the DMI prevents the formation of other topological excitations, in sharp contrast to the variety of two-dimensional skyrmions that appear in other models~\cite{Piette}. 
The presence of only one type of excitation defines the approach for data encoding, which is based on the quantization of distances between adjacent excitations on a track.
However, skyrmions are highly movable, interacting objects that can drift as a result of of thermal fluctuations, making it difficult to maintain their distribution along a track.
The fabrication of arrays of artificial pinning centers on the nanoscale to solve this problem is a serious challenge that will also lead to higher costs.
In a system with surface/interface-induced DMI, a solution is the fabrication of a nanostripe with a special profile that results in the location of skyrmions in two parallel channels~\cite{Muller}.

Here, we  propose a powerful alternative approach, which does not require fixed distances between  movable bit carriers because the data stream is encoded in a single chain that is composed of two distinct particle-like states.

The class of magnetic materials, in which there is a significant contribution from the DMI and in which magnetic skyrmions have been observed is based on non-centrosymmetric crystals~\cite{Kanazawa2017} such as B20-type MnSi~\cite{Yu_15}, FeGe~\cite{Yu_11}, Fe$_{1-x}$Co$_x$Si~\cite{Yu_10}, Mn$_{1-x}$Fe$_x$Ge \cite{Shibata_13} and high temperature $\beta$-Mn-type Co-Zn-Mn alloys~\cite{Tokunaga_15}.
In such materials, a competition between ordinary Heisenberg exchange interactions and the DMI results in a spin spiral ground state.
The equilibrium period of such a spin spiral state, $L_\textrm{D}\!=\!4\pi\mathcal{A}/\mathcal{D}$, depends on an interplay between exchange stiffness, $\mathcal{A}$, and the DMI constant, $\mathcal{D}$~\cite{helix}.
In an applied magnetic field, the behavior of such a system depends strongly on the dimensionality of the sample.
In a bulk sample, the spin spiral state usually appears as a multidomain state, with the \textbf{k}-vector of the spiral pointing in different crystallographic directions in different domains.
%
In the presence of an applied magnetic field, such a multidomain spiral state transforms into a monodomain conical state, with the magnetization tilted towards the direction of the applied magnetic field, $\textbf{H}_\mathrm{ext}$ and with $\mathbf{k}\parallel\mathbf{H}_\mathrm{ext}$.
The conical state persists as the lowest energy state over the entire range of applied magnetic fields up to a critical value of $H_\textrm{D}\!=\!\mathcal{D}^2/(2\mathcal{A}M_\textrm{s})$.
For $H_\mathrm{ext}\!\geqslant\!H_\textrm{D}$, it converges to the saturated ferromagnetic state.

In a thin film of a chiral magnet of thickness $L\sim L_\textrm{D}$, the energy of magnetic skyrmions is lower than that of the conical phase over a wide range of applied magnetic  fields~\cite{Rybakov_13} and skyrmions appear in the form of a hexagonal lattice~\cite{Yu_10}.
It has been shown both experimentally~\cite{Yu_11} and theoretically~\cite{Rybakov_13} that the range of skyrmion existence depends on the film thickness.
For example, there is a critical thickness above which the skyrmions in such a system appear only as a metastable state.

Although it has been assumed that the magnetic skrymion is a unique localized spin texture in chiral magnets, it was recently proposed that 
for film thicknesses that are larger than the equilibrium period,
$L\!\gtrsim\!L_\textrm{D}$, and magnetic fields, $H_\mathrm{ext}\!<\!H_\textrm{D}$, magnetic skyrmions may coexist with another type of localized particle-like object - the chiral bobber (ChB) (Fig.~\ref{Fig-intro}a).
The proposed mechanism of ChB stabilization and its range of existence have been discussed theoretically elsewhere.~\cite{Rybakov_15,Rybakov_16}.
Significantly, there is predicted to be a certain range of film thicknesses and applied fields, over which the energies of isolated skyrmion tubes (SkTs) and ChBs as excited states are comparable or even equal, suggesting that the two states may coexist.
Since the energy barriers that protect skyrmions and ChBs are also comparable~\cite{Rybakov_15}, these two objects are potentially ideal candidates for use as ``1'' and ``0'' bit carriers (Fig.~\ref{Fig-intro}a). However, ChBs have not previously been observed experimentally.

As a ChB occupies a significantly smaller volume in a film than a SkT, the two states can in principle be distinguished from one another by their net magnetization, electronic transport properties or stray fields.
Here, we compare the results of micromagnetic simulations (Fig.~\ref{Fig-intro}b-d) with quantitative experimental measurements performed using the technique of off-axis electron holography (EH) \cite{RDB09} in the transmission electron microscope (TEM). EH provides high spatial resolution  measurements of the phase shift originating from the interaction between an incident electron beam and the projected in-plane component of the magnetic induction within and around the sample.

When the applied field is too strong, $H_\mathrm{ext}\!\gtrsim\!400$~mT, ChBs are predicted to shrink and the resulting signal to be weak (Fig.~\ref{Fig-intro}b,~e).
Above this field, the ChBs collapse, while SkTs remain stable up to $H_\mathrm{ext}\!\sim\!500$ mT. 
There is an optimal range of applied magnetic fields between 250 and 350~mT, in which the signals from ChBs and SkTs can be distinguished reliably (Fig.~\ref{Fig-intro}c,~f).
For very low fields, $H_\mathrm{ext}\!\lesssim\!200$ mT, it is difficult to do so because of strong distortions of the SkTs (Fig.~\ref{Fig-intro}d,~g).
Equilibrium distances between particles in Fig.~\ref{Fig-intro}b-d, after energy minimization, indicate that SkTs and ChBs are coupled via internal forces and have a tendency to condense into compact chains or clusters in low values of applied field.

\begin{figure*}
\centering
\includegraphics[width=18cm]{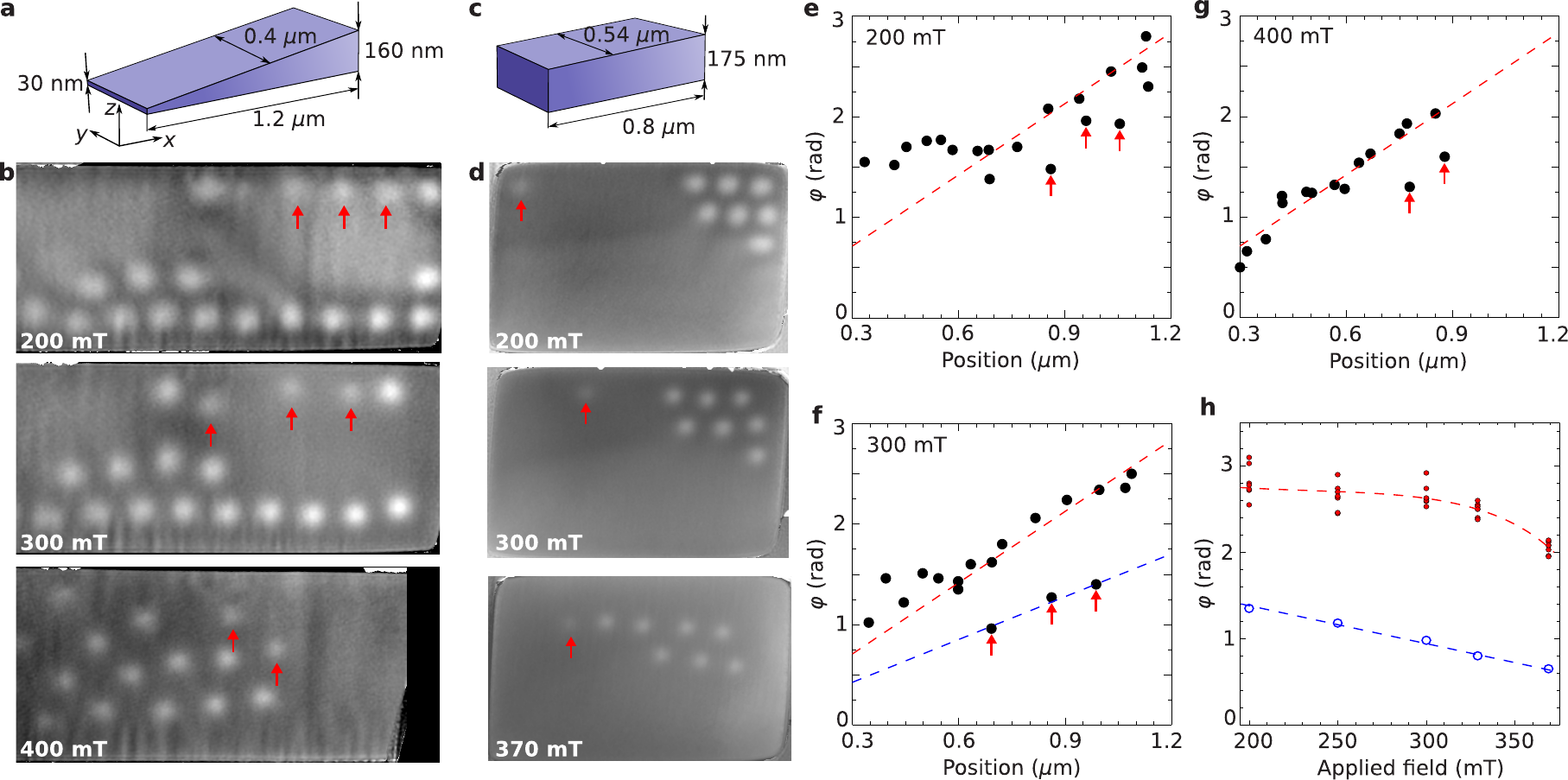}
\caption{\small \textbf{Experimental evidence for chiral magnetic bobbers.}
\textbf{a},~\textbf{b},~Size and geometry of a wedge-shaped \textit{B}20 FeGe sample and magnetic phase shift images recorded in the presence of different magnetic fields applied perpendicular to the sample plane, respectively.
\textbf{c},~\textbf{d},~Size and geometry of a fixed thickness \textit{B}20 FeGe sample and magnetic phase shift images recorded in the presence of different magnetic fields applied perpendicular to the sample plane, respectively.
All of the images shown in \textbf{b},~\textbf{d} were recorded at a temperature $T\!=\!95$~K, in order to avoid thermal drift of the magnetic structures.
\textbf{e-g},~Phase shifts in the wedge-shaped sample described in \textbf{a}, recorded in the presence of external magnetic fields of 200, 300 and 400~mT, respectively.
The points correspond to the values of the phase shift signal in the centers of the bright spots in \textbf{b} and are presented as function of the positions of the spots from the left (thin) edge of the wedge-shaped sample.
THe dotted lines mark trend lines of the phase shift signal for SkTs (red) and ChBs (blue). The error in each phase measurement is estimated to be $\pm$0.05 rad. 
The red arrows indicate objects with weak contrast that are identified as ChBs.
\textbf{h},~Dependence of phase shift measured for different applied magnetic fields in the sample of fixed thickness shown in \textbf{c}. Solid (red) circles correspond to SkTs, while open (blue) circles correspond to ChBs.
}
\label{Fig-experiment}
\end{figure*}

In order to identify an appropriate range of magnetic fields and film thicknesses, in which  SkTs and ChBs are expected to coexist, we performed EH experiments (Fig.~\ref{Fig-experiment}) using two different specimen geometries of \textit{B}20-type FeGe.
Due to the relatively high magnetic moment of $\mu_\mathrm{Fe}\sim1\mu_\mathrm{B}$, the stray fields, which depend on the shape and size of the sample, make a significant contribution to the total magnetic energy~\cite{Shibata_17}.
As it is difficult to estimate the optimal thickness for ChB formation in advance, we first studied a wedge-shaped specimen (Fig.~\ref{Fig-experiment}a), whose thickness varied from 30 to 160~nm.
Fig.~\ref{Fig-experiment}b shows the magnetic phase shift recorded experimentally from a wedge-shaped sample in magnetic fields of 200, 300 and 400~mT, which were applied perpendicular to its plane.
The phase shift images show contrast that is typical for magnetic skyrmions, with bright disks of nearly identical intensity.
However, quantitative analysis of the phase shift intensities relative to the background conical or saturated ferromagnetic state reveals the presence of two distinct types of object, whose signals have different intensities.
This difference is most evident at an intermediate applied magnetic field of 300~mT (Fig.~\ref{Fig-experiment}f). 
The value of the phase shift signal in the center of each bright spot in Fig.~\ref{Fig-experiment}b is plotted in Fig.~\ref{Fig-experiment}e-g as a function of distance from the left (thin) edge of the sample.
According to the micromagnetic calculations described above, objects that have a higher phase shift (corresponding to data points along the red dashed trend line) are identified as SkTs, while those exhibiting weaker contrast (corresponding to data points along the blue dashed trend line and marked by red arrows) are identified as ChBs.

In order to obtain a state with multiple SkTs and ChBs (Fig.~\ref{Fig-experiment}b), we performed a series of heating-cooling cycles for different maximum temperatures and different fixed values of applied magnetic field, typically $T_\textrm{max}\!\approx\!0.9\,T_\textrm{c}$ (for FeGe $T_\textrm{c}\!=\!278$\,K).
Between successive heating-cooling cycles, we also performed an additional cycle of magnetization reversal by varying the absolute value and direction of the applied magnetic field.
In order to avoid thermal drift of the objects, off-axis electron holograms were recorded at a reduced temperature of $T\!=\!95$\,K.

At $H_\mathrm{ext}\!=\!200$\,mT, the ChBs and SkTs are observed to form chain-like clusters along the sample edges with approximately equal distances between them, confirming similar interactions between the ChBs, SkTs and edges according to a Lennard-Jones type potential, as has been found for SkTs~\cite{Leonov}.
In Fig.~\ref{Fig-experiment}e, it is possible to see the different contrast of the objects. However, in agreement with results of the micromagnetic simulations, which predict strong distortions of the shapes of SkTs or ChBs in low magnetic fields, it is difficult to distinguish them directly from the recorded images.

With increasing applied magnetic field, the equilibrium distance between the particles increases and the contrast in the phase shift approaches the predicted trend lines (Fig.~\ref{Fig-experiment}f). 
For $H_\mathrm{ext}\!=\!300$\,mT, the contrast of the ChBs is approximately twice as weak as the signal from the the SkTs, in agreement with theory.
As the applied magnetic field is increased further, the particles form a cluster in the central part of the sample.
The absence of a ChB at 400~mT (Fig.~\ref{Fig-experiment}b,~g) indicates that at $T\!=\!95$\,K the field is close to the critical value for the existence of ChBs.

In order to identify the existence and stability of ChBs in a sample of nearly constant thickness, we performed an additional experiment (Fig.~\ref{Fig-experiment}c).  
By following the procedure described above, we performed several magnetization reversal cycles, as well as several heating-cooling cycles at a fixed magnetic field, in order to observe the spontaneous nucleation of ChBs in this sample.
The EH phase images shown in Fig.~\ref{Fig-experiment}d illustrate a cluster composed of a few SkTs, as well as a single ChB, which is marked by a red arrow.
Fig.~\ref{Fig-experiment}h shows the dependence of the signal intensity for SkTs and ChBs measured in this sample at different values of applied magnetic field. 
In good agreement with the theoretical model, the intensities of the ChBs and SkTs decrease gradually with increasing applied magnetic field.
Fig.~\ref{Fig-experiment}f,~h shows that the difference in signal intensity between ChBs and SkTs increases with both film thickness and applied magnetic field.
This dependence allows the ratio of signals that is required for reliable data reading and subsequent signal decoding to be maintained.

It should be noted that the present EH measurements do not allow reconstruction of the three dimensional (3D) magnetic configuration and only provide information about the in-plane magnetic induction averaged through the thickness of the film.
As a result, it is not possible to determine whether the ChBs are located at the top or the bottom surface of the sample, or whether they form a coupled state with two ChBs located at both surfaces, as predicted theoretically~\cite{Rybakov_13,Rybakov_15}.
We expect that the increase in the signal of the ChBs in Fig.~\ref{Fig-experiment}g can be explained by the appearance of such a coupled state.
Taking into account the facts that (i)~the objects with abnormally weak contrast appear only above a certain film thickness, (ii)~they exhibit a field dependence of the contrast according to a single trend line, (iii)~they are pinning-free movable objects and (iv)~the good agreement with theoretical predictions, we conclude that the coexistence of ChBs and skyrmions is confirmed by our experimental results.

In conclusion, our work provides the direct observation of theoretically predicted particlelike objects that coexist together with magnetic skyrmions in thin films of cubic chiral magnets. The presence of two types of localized movable object allows us to introduce a new concept of magnetic data storage, which relies on these two objects taking on the role of binary states of data bits. Their coherent motion is expected to be stable because of cohesion effects due to interparticle interactions. A higher data density can be achieved in comparison to the existing skyrmion-based racetrack memory concept.
Future experimental studies will focus on an efficient approach for ChB nucleation and mobility in the presence of an electric current, as well as studies in other materials and using other experimental techniques.

\subsection*{Methods}
\subsection*{Micromagnetic simulations}
Initial micromagnetic simulations for a simplified model without dipole-dipole interaction, where the phenomenon of merging of SkTs and ChBs into chains or clusters was found, were performed using software described in~ \onlinecite{Rybakov_15}. 
Final micromagnetic simulations were performed using MuMax$^3$~\cite{MuMax3} for the following energy functional:
\begin{eqnarray}
\mathcal{E} \!= \!\int\limits_{V_\mathrm{s}}  
\left\lbrace  
\mathcal{E}_\mathrm{ex}+
\mathcal{E}_\mathrm{DMI}+
\mathcal{E}_\mathrm{Z}+
\mathcal{E}_\mathrm{d}
\right\rbrace 
\mathrm{d}{\bf r},\label{E_tot} \\
\mathcal{E}_\mathrm{ex}=\mathcal{A} \left( {{\partial}_x\mathbf{m}}^2 + {{\partial}_y\mathbf{m}}^2+ {{\partial}_z\mathbf{m}}^2   \right), \nonumber \\
\mathcal{E}_\mathrm{DMI}=\mathcal{D}\, \mathbf{m}\cdot [\mathbf{\nabla} \times \mathbf{m}   ], \nonumber \\
\mathcal{E}_\mathrm{Z}= - M_\mathrm{s}\, \mathbf{H}_\mathrm{ext}\cdot \mathbf{m},\nonumber \\
\mathcal{E}_\mathrm{d}= -\frac{1}{2} M_\mathrm{s}\, \mathbf{H}_\mathrm{d} \cdot \mathbf{m},\nonumber 
\end{eqnarray} 
where  $\mathbf{m} \equiv \mathbf{m}(x,y,z)$ is a normalized ($|\mathbf{m}|=1$) continuous vector field, $\mathcal{A}$ and $\mathcal{D}$ are the micromagnetic constants for exchange and DMI, respectively, $M_\mathrm{s}$ is the magnetization of the material, $\mathbf{H}_\mathrm{ext}$ is the applied magnetic field and  $\mathbf{H}_\mathrm{d}$ is the demagnetizing field generated by the divergence of the magnetization in the volume of the sample and at its edges. The integration in Eq.~\ref{E_tot} extends over the entire volume of the sample, $V_\mathrm{s}$. 
In our calculations, we assume the following material parameters: $\mathcal{A}=4.75$ pJm$^{-1}$, $\mathcal{D}=0.852718$ mJm$^{-2}$ and $M_\mathrm{s}=384$ kAm$^{-1}$, resulting in an equilibrium period of the spin spiral of $L_\mathrm{D}=70$~nm, which is a typical value for FeGe.
The mesh size of the simulated domain is 528~$\times$~82~$\times$~64 cells, where each cell is a cube with an edge length of 2~nm.
In order to achieve a state with alternating skyrmions and ChBs, as shown in Fig.~\ref{Fig-intro}b-d, we started the simulations from a homogeneously magnetized state, with imposed SkTs penetrating the entire thickness of the nanostripe or with tubes that end up in the middle plane of the sample. 
After energy minimization using a conjugate gradient method, the spin structure was exported to the Spirit code~\cite{Spirit} for visualization.

It should be noted that the presence of Bloch point magnetic singularities does not lead to a divergence in total energy \cite{Doring}. The dynamics and equilibrium configurations of the system can still be calculated using a micromagnetic approach, as has been shown for permalloy disks \cite{Thiaville2003} and thin films \cite{Zverev2013}.

\subsection*{Off-axis electron holography in the TEM}
TEM specimens of controlled thickness and dimensions were prepared from a single crystal of \textit{B}20-type FeGe using a focused ion beam (FIB) workstation and a lift-out method ~\cite{Du_15}.  
Off-axis electron holograms were recorded using an electrostatic  biprism positioned in a conjugated image plane in an FEI Titan 60-300 TEM operated at 300~kV. The microscope was operated in aberration-corrected Lorentz mode with the sample in magnetic-field-free conditions. The conventional microscope objective lens was then used to apply chosen vertical (out-of-plane) magnetic fields of between -0.15 and 1.5~T, which were pre-calibrated using a Hall probe. A liquid nitrogen specimen holder (Gatan model 636) was used to vary the sample temperature between 95 and 380~K. Images and holograms were recorded using a 2k~$\times$~2k charge-coupled device (CCD) camera (Gatan Ultrascan). 30 object holograms and 30 vacuum reference holograms, with a 6~s exposure time for each hologram, were recorded to improve the signal-to-noise ratio. The off-axis electron holograms were analyzed using a standard fast Fourier transform (FFT) algorithm in Holoworks software (Gatan). In order to remove the mean inner potential (MIP) contribution from the total phase shift, phase images were recorded both at low temperature and at room temperature, aligned and subtracted from each other on the assumption that the MIP is the same and there are no signficant changes in specimen charging between the two holograms.

\section*{Acknowledgments}
This work was supported by the Natural Science Foundation of China, Grant No. 51622105, 11474290, the Key Research Program of Frontier Sciences, CAS, Grant No. QYZDB-SSW-SLH009 and the Youth Innovation Promotion Association CAS No. 2015267. F.Z. and R.E.D-B acknowledge the European Union for funding through the Marie Curie Initial Training Network SIMDALEE2. The research leading to these results received funding from the European Research Council under the European Union's Seventh Framework Programme (FP7/2007-2013)/ ERC grant agreement number 320832.

\section*{Author contributions}
F.N.R. proposed the concept and, together with A.B.B., performed preliminary simulations.
F.Z., D.S., S.W. and H.D. performed the experiments.
H.D., A.K. and R.E.D-B. supervised and designed the experiments.
N.S.K performed micromagnetic simulations and prepared the initial version of the manuscript.
J.C. calculated theoretical phase shift images.
M.T., S.B. and R.E.D-B. edited the manuscript.
All of the authors discussed the results and contributed to the preparation of the manuscript.

\end{document}